\documentclass[10pt,conference]{IEEEtran} 
\IEEEoverridecommandlockouts
\usepackage{cite}
\usepackage{amsmath,amssymb,amsfonts}
\usepackage{textcomp}
\usepackage{xcolor}
\usepackage{multirow}

\usepackage{wrapfig}

\usepackage{pifont}
\usepackage{xspace}
\usepackage{soul}
\usepackage{changepage,threeparttable} 
\usepackage{marginnote}
\usepackage{listings}

\usepackage{color}
\usepackage{graphicx}
\usepackage{subfigure}
\usepackage{epstopdf}
\usepackage{hyperref}

\AtBeginDocument{%
  \providecommand\BibTeX{{%
    \normalfont B\kern-0.5em{\scshape i\kern-0.25em b}\kern-0.8em\TeX}}}

\begin{document}

\onecolumn
  
\newenvironment{project}[1]
{\par
 \bigskip
 \begin{wrapfigure}{l}[0pt]{1in}
 \vspace{-15pt}
 \includegraphics[width=1in,clip,keepaspectratio]{#1}
 \vspace{-25pt}
 \end{wrapfigure}
 \footnotesize \noindent}
{\par\bigskip}

\newenvironment{bio}[1]
{\par
 \bigskip
 \begin{wrapfigure}{l}[0pt]{0.5in}
 \vspace{-15pt}
 \includegraphics[width=0.5in,clip,keepaspectratio]{#1}
 \vspace{-25pt}
 \end{wrapfigure}
 \footnotesize \noindent}
{\par\bigskip}

\begin{figure}
    \centering
    \includegraphics[width=.3\textwidth]{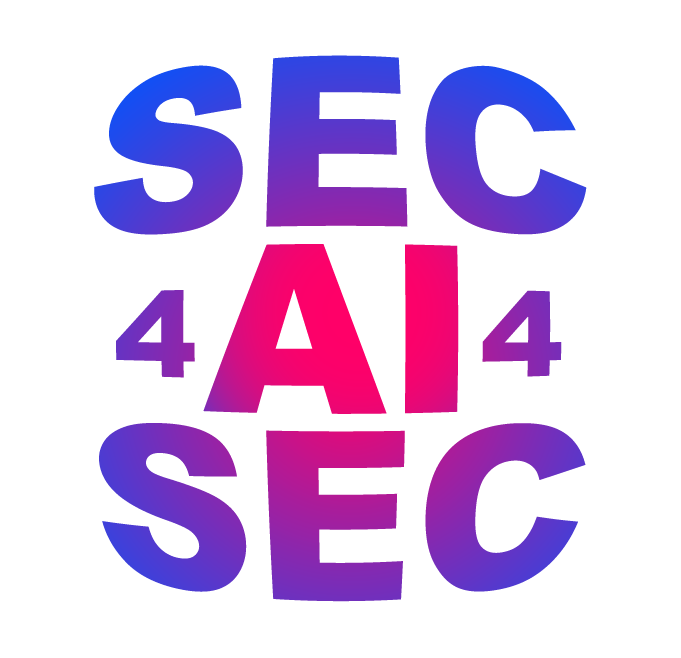}
\end{figure}

\vspace{2\baselineskip}


\vspace{2\baselineskip}

\begin{center}
{\huge \textbf{Using ML filters to help automated vulnerability
repairs: when it helps and when it doesn’t}}
\end{center}

\vspace{\baselineskip}

{\large
Authors:
\begin{itemize}
    \item[]\textbf{Maria Camporese}, University of Trento (Italy)
    \item[]\textbf{Fabio Massacci}, University of Trento (Italy), Vrije Universiteit Amsterdam (The Netherlands)
\end{itemize}
}

\vfill

\begin{figure}[h]
\vspace{-\baselineskip}
\includegraphics[height=3cm]{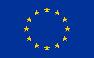}
~\includegraphics[height=3cm]{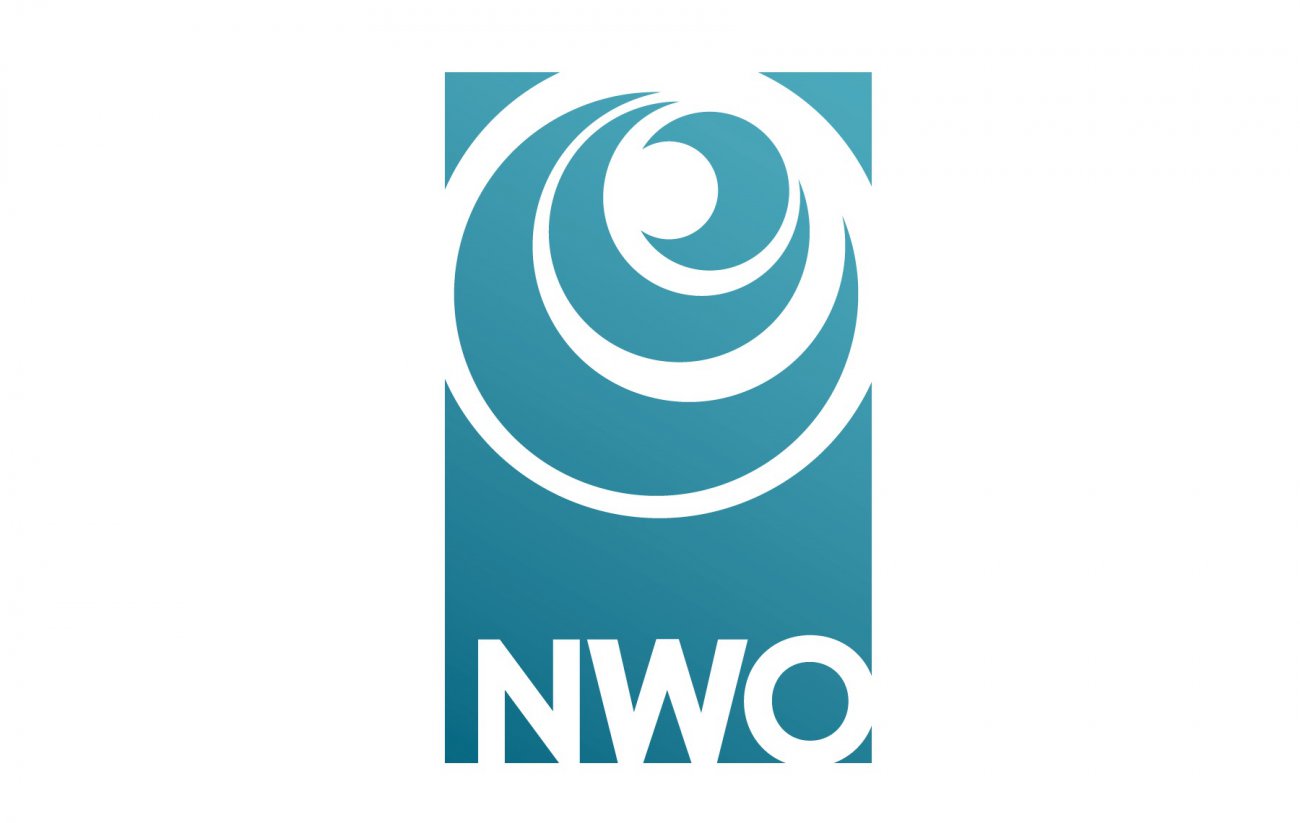}
~\includegraphics[height=3cm]{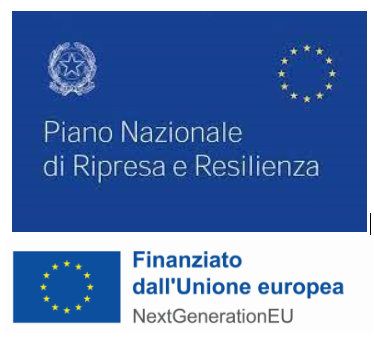}

\end{figure}

\noindent This work has been partly supported by the European
Union (EU) under Horizon Europe grant n
. 101120393
(Sec4AI4Sec), by the Nederlandse Organisatie
voor Wetenschappelijk Onderzoek (NWO) under grant n.
KIC1.VE01.20.004 (HEWSTI), and by the Italian Ministry of University and Research (MUR), under the P.N.R.R. – NextGenerationEU grant
n. PE00000014 (SERICS). This paper reflects only the author's view and the funders are not responsible for any use that may be made of the information contained therein.

\clearpage
\onecolumn
\begin{project}{logos/EU-Logo.png}
\textbf{Cybersecurity for AI-Augmented Systems (Sec4AI4Sec)} . As artificial intelligence (AI) becomes omnipresent, even integrated within secure software development, the safety of digital infrastructures requires new technologies and new methodologies, as highlighted in the EU Strategic Plan 2021-2024. To achieve this goal, the EU-funded Sec4AI4Sec project will develop advanced security-by-design testing and assurance techniques tailored for AI-augmented systems. These systems can democratise security expertise, enabling intelligent, automated secure coding and testing while simultaneously lowering development costs and improving software quality. However, they also introduce unique security challenges, particularly concerning fairness and explainability. Sec4AI4Sec is at the forefront of the move to tackle these challenges with a comprehensive approach, embodying the vision of better security for AI and better AI for security. More information at \textbf{\url{https://sec4ai4sec.eu}}.
\end{project}

\begin{project}{logos/NWO-logo.jpg}
\textbf{Hybrid Explainable Workflows for Security and Threat Intelligence (HEWSTI)} In research into threats to safety and security, people and AI collaborate to obtain actionable intelligence. Their sources and methods often have significant uncertainties and biases. Experts are aware of these limitations, but lack the formal means to handle these uncertainties in their daily work. This project will invent a ‘metadata of uncertainty’ for threat intelligence (in both machine-readable and also human-interpretable forms) and validate it empirically. Intelligence agencies will then be able to explicitly consider the trade-off between the accuracy, proportionality, privacy, and cost-effectiveness of investigations. This will contribute towards the responsible use of AI to create a safer, more secure society.
\end{project}

\begin{project}{logos/pnrr-next_generation-logo.png}
\textbf{In searCh Of eVidence of stEalth cybeR Threats
(COVERT)} AT 3 aims to analyze emerging attack methodologies and develop advanced methods for detecting attacks and identifying guidelines for designing IT systems that ensure reduced vulnerability to new attack categories. The detailed objectives can be divided into four macro categories: (i) Development of advanced tools for analyzing malware and software aimed at identifying vulnerabilities that could be exploited by malware; (ii) Development of tools for analyzing network traffic to identify communications related to ongoing attacks; (iii) Development of machine learning systems that are robust to attacks and through which it is possible to extract knowledge aimed at creating more advanced tools for timely analysis and early identification of attacks; (iv) Analysis of the "human factors" involved in an attack with the development of tools for analyzing and correlating information from OSINT (open sources intelligence) and for the defense and prevention of attacks based on social engineering techniques.
\end{project}

\vfill

\begin{bio}{photos/camporese}
\textbf{Maria Camporese} (MSc 2022) is a PhD student at the University of Trento, Italy. Her research interests include security and machine learning. Contact her at \emph{maria.camporese@unitn.it}.
\end{bio}

\begin{bio}{photos/massacci}
\textbf{Fabio Massacci} (Phd 1997) is a professor at the University of 
Trento, Italy, and Vrije Universiteit Amsterdam, Fabio Massacci is a professor at the University of Trento, Trento, Italy, and Vrije Universiteit Amsterdam, 1081 HV Amsterdam, The Netherlands. His research interests include empirical methods for the cybersecurity of sociotechnical systems. For his work on security and trust in sociotechnical systems, he received the Ten Year Most Influential Paper Award at the 2015 IEEE International Requirements Engineering Conference. He is named co-author of CVSS v4. He leads the Horizon Europe Sec4AI4Sec project and the Dutch National Project HEWSTI.  He is past chair of the Security and Defense Group of the Society for Risk Analysis, and IEEE CertifAIEd Lead Assessor. Contact him at \emph{fabio.massacci@ieee.org}.
\end{bio}

How to cite this paper:
\begin{itemize}
    \item Camporese, M. and Massacci, F. Using ML filters to help automated vulnerability repairs: when it helps and when it doesn’t. \emph{Proceedings of the International Conference on Software Engineering - New and Emerging Results (ICSE-NIER 2025)}. IEEE Press.
\end{itemize}

License:
\begin{itemize}
\item This article is made available with a perpetual, non-exclusive, non-commercial license to distribute.
\end{itemize}

\clearpage

\twocolumn
\newcommand{\cmark}{\ding{51}}%
\newcommand{\xmark}{\ding{55}}%

\definecolor{BoxGray}{gray}{0.93}
\long\def\finding#1{
\par\noindent\colorbox{BoxGray}{\fbox{\parbox{0.95\columnwidth}{\emph{#1}}}}\par}

\long\def\mc#1{\marginpar[Maria]{Maria}\begin{color}{teal}#1\end{color}}
\long\def\fm#1{\marginpar[Fabio]{Fabio}\begin{color}{orange}#1\end{color}}

\def\prevalence{\ensuremath{\pi}\xspace}
\def\ppv#1{\ensuremath{P_{#1}}\xspace}
\def\time#1{\ensuremath{\tau_{#1}}\xspace}
\def\tpr#1{\ensuremath{R_{#1}}\xspace}
\def\fpr#1{\ensuremath{Far_{#1}}\xspace}
\def\filter{\ensuremath{{\texttt{V}}}\xspace}
\def\model{\ensuremath{{\texttt{M}}}\xspace}
\def\negmodel{\ensuremath{{\texttt{MVD}}}\xspace}
\def\totalpatches{\ensuremath{n}\xspace}

\newcommand{\ea}{{et al.}\xspace}
\newcommand{\thead}[1]{\textbf{\textit{#1}}}

\graphicspath{img/}



\title{Using ML filters to help automated vulnerability repairs: when it helps and when it doesn't}

\author{\IEEEauthorblockN{Maria Camporese}
\IEEEauthorblockA{
\textit{University of Trento}, IT \\
maria.camporese@unitn.it}
\and
\IEEEauthorblockN{Fabio Massacci}
\IEEEauthorblockA{\textit{University of Trento}, IT}
\textit{Vrije Universiteit Amsterdam}, NL\\
fabio.massacci@ieee.org}




\maketitle

\pagestyle{plain}
\begin{abstract}
[Context:] The acceptance of candidate patches in automated program repair has been typically based on testing oracles. Testing requires typically a costly process of building the application while ML models can be used to quickly classify patches, thus allowing more candidate patches to be generated in a positive feedback loop.
[Problem:] If the model predictions are unreliable (as in vulnerability detection) they can hardly replace the more reliable oracles based on testing. 
[New Idea:] We propose to use an ML model as a preliminary filter of candidate patches which is put in front of a traditional filter based on testing. 
[Preliminary Results:] We identify some theoretical bounds on the precision and recall of the ML algorithm that makes such operation meaningful in practice. With these bounds and the results published in the literature, we calculate how fast some of state-of-the-art vulnerability detectors must be to be more effective over a traditional AVR pipeline such as APR4Vuln based just on testing. 
\end{abstract}


\section{Introduction and Problem Statement}
The ultimate goal of Automated Program Repair (APR) pipelines \cite{shen2020survey} is identify a faulty code fragment, generate a patch, validate it, and ultimately propose it to a human developer, who will either discard or accept it. In this respect, two dimensions are important: speed of patch generation \cite{winter2022developers} and a (small) number of quality patches surviving the process \cite{noller2022trust}. Williams et al. \cite{williams2024user} showed that by improving the way, time, and context in which APR patches are suggested significantly increased their adoption rate at Bloomberg.

Many techniques have been used for fault localization (from  model-checking  \cite{charalambous2023new} to static analyzers \cite{jasz2022end}) and patch generation (from pattern-based mechanisms \cite{jasz2022end} to Large Language Models \cite{charalambous2023new,jin2023inferfix}), but testing remains the most widespread practice for patch validation \cite{avgerinos2014automatic,pinconschi2021comparative,bui2024apr4vul} as static analysis has many false positives \cite{hegedHus2022static}. Even having passed all end-to-end tests is not enough to ensure the patch is correct \cite{liu2019icseeval} and for vulnerability testing this is even harder \cite{bui2024apr4vul}. Table~\ref{tab:validation:criteria}  summarizes the pros and cons of different validation criteria.
\begin{table*}

\centering
\caption{Pros and cons of patch validation criteria}
\begin{minipage}{\linewidth}{\itshape \footnotesize

\vspace{0.1cm}
}\end{minipage}
\renewcommand{\arraystretch}{1,2}
\begin{tabular}{p{0.15\textwidth}p{0.2\textwidth}p{0.55\textwidth}}
    \hline 
    \thead{Validation criterion}  &
    \thead{Pros} &
    \thead{Cons}\\ \hline

    \textbf{Test cases} \newline
    \cite{wu2023effective} \cite{pinconschi2021comparative} \cite{bui2024apr4vul} \cite{jasz2022end}   & Execution is automatic and grants perfect recall & Generating complete and reliable tests require developer time and its possible for already-known code faults only. Patches could overfit tests\\ \hline

    \textbf{Human-validation} \newline
    \cite{wu2023effective} \cite{pinconschi2021comparative} \cite{bui2024apr4vul} \cite{jasz2022end}  & Reliable method to recognize variants of correct solutions & Slow, costly, subject to human errors
    \\ \hline
    
    \textbf{Static Code Analysis (SCA) tools} \cite{jasz2022end}
     & 
    Automatic, not require information only available in hindsight 
    & SCA suffer from a high false positive rate and patches could overfit on warnings\\ \hline
    
    \textbf{Perfect match} \newline
    \cite{jin2023inferfix} \cite{fu2022vulrepair}  & Automatic and grants perfect precision
    & As developer-generated fixes are only available in hindsight, it cannot be applied new vulnerabilities. It also excludes all solutions semantically equivalent to the ground truth. 
    \\ \hline
\end{tabular}
\label{tab:validation:criteria}
\end{table*}

Recently, Machine Learning (ML) has been also proposed for patch generation and validation \cite{fu2022vulrepair,chi2022seqtrans}. In constrast to testing (or static analysis), ML does not requires to build and run the application, a costly process in terms of time and required infrastructure.   Unfortunately, an ML model only can only predict the similarity with patches that it has seen in the past. ML filtered patches might pass because they overfit rather than because they provide an actual solution \cite{le2018overfitting}. 

In the realm of automated vulnerability repair (AVR), replacing testing with ML is particularly challenging. Allegedly good vulnerability detectors are retrospectively discovered to be poor by independent analysis \cite{chakraborty2021deep}. A recent study \cite{wen2023vulnerability} showed that when performance was calculated on different datasets precision dropped to 50\% or even lower. Independent stduies \cite{bui2024apr4vul,papotti2024acceptance} of the ML-based system for vulnerability repairs  SeqTrans \cite{chi2022seqtrans} found that almost all patches who survived end-to-end testing were semantically incorrect.

So if ML models are too fragile for replacing testing in AVR
can we still benefit from their ability to quickly classify (albeit possibly incorrectly) a candidate patch?




\section{Research Idea}
\label{sec:research_challenges}

\finding{
\textbf{\textit{Overarching Hypothesis.}} 
Breaking down the patch validation phase into steps ordered by their computational complexity and including ML-driven "fail fast" steps can improve the efficacy and efficiency of APR pipelines.}

Figure \ref{fig:ai-augmented-system} illustrates the idea on how the ML-filter can be used in practice. Patches could be pre-screened by a ML model before undergoing traditional testing. This type of filter has been first proposed to select `human-like patches' \cite{bader2019getafix}. We propose to generalize it to quickly exclude incorrect patches, while sounder but slower validation steps could provide a final assurance for promising patches. We see two advantages: 
\begin{figure*}[t]
\subfigure[Patch validation in the APR process.]{\includegraphics[width=0.5\textwidth]{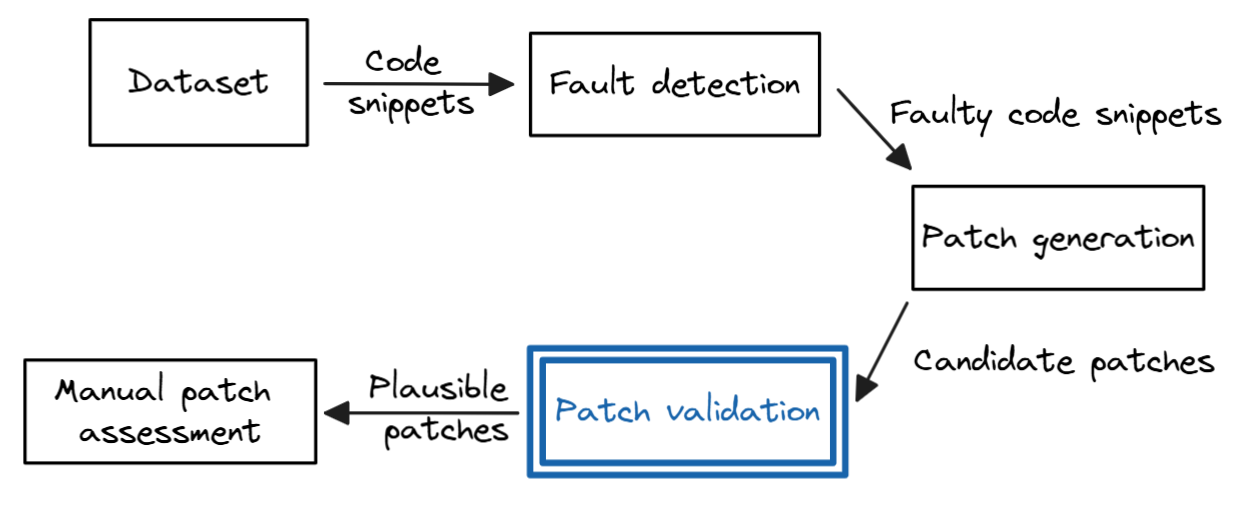}\label{fig:1a}}
\subfigure[Pre-screening a validation filter with a ML model.] {\includegraphics[width=0.50\textwidth]{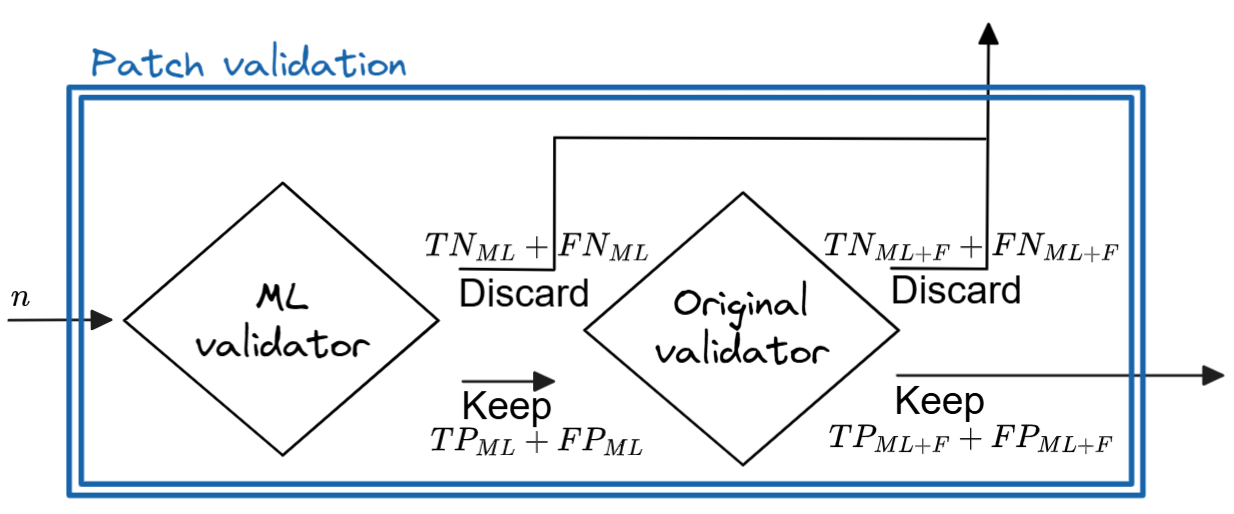}\label{fig:1b}}
  \begin{minipage}{\linewidth}
\footnotesize \textit{Left.} While most of the APR pipelines proposed in the literature adopt a single-step approach for validation, we decompose it in a progressive process in which a first filter aims to exclude most of the unpromising repair attempts before they reach the second, more expensive validation step. 

\textit{Right.} We consider the case in which an ML model that performs binary classification is used to pre-screen patches before a given validation filter. In Section \ref{sec:execution_plan}, we use the model, filter and data distribution properties to provide an estimate of whether the model introduction could improve the validation process.

\vspace*{-0.1\baselineskip}
\end{minipage}
  \caption{Enhancing validation with an ML validator}
  \label{fig:ai-augmented-system}
\end{figure*}

\paragraph{Time efficiency} First, quickly excluding a significant number of bad patches could save time on the overall validation process. Querying an ML model is allegedly quick and a model can be pushed to consider some misclassification errors more than others. 
A custom loss function could weight differently discarded good patches vs bad patches kept in the process, as the latter only clog the validation process. 

\paragraph{Validation effectiveness} A second potential advantage of using independent filters could be improving the efficacy of patch generation and validation. 
If querying the ML model is fast enough, part of the time it saves could be used to generate more candidate patches. Including enough attempts could preserve the throughput of good patches even if the model were to exclude some of them.

However, just deploying a `fail fast' ML screener may not guarantee that the overall pipeline will offer more good repairs to the human developer. Before running a costly data collection and training effort, are there some \emph{general mathematical conditions} under which the insertion of a pre-screening ML model before a particular validation filter is convenient for the overall performance of APR pipelines?

\section{Is an ML pre-screening always convenient?}
\label{sec:execution_plan}
\subsection{Key Specification Parameters}
We want to use only the indicators that could be found in the specs of the ML model or the existing validation filters. Given \totalpatches candidate patches considered in the given time-frame, and the expected prevalence rate \prevalence of APR generated good patches over the total number of patches \totalpatches, we only require the precision \ppv {i} and the recall \tpr {i} of the $i$-th step of the pipeline. We also need the time \time i\ of the $i$-th filter uses to evaluate a single patch. For example, $\time\model$ denotes the time the ML model \model uses to make a single prediction, while $\tpr \filter$ is the Recall of the traditional validator \filter used to identify good patches.

\subsection{Bounding the effectiveness}
The key intuition is that if the ML model is able to `fail fast' incorrect patches, part of the time thus saved could be used to generate and test more candidates.
A larger pool of candidate patches could increase the number of positive patches surviving the overall pipeline and thus providing a more effective solution to the human end user. Due to lack of space, we do not consider the time for patch generation: if an ML model is not convenient when patch generation is free, it would also not be useful when it costs. 

This intuition can be captured by two formal requirements: (i) the number of good patches surviving the traditional filter \filter processing the initial number of patches $n$ is at least equal to the number of good patches surviving the augmented pipeline (with bodth the model \model and the validator \filter) processing the larger number of patches $n+\Delta n$, and (ii) the pipeline takes less time, even if it has to filter also the additional patches. 
\begin{eqnarray}
    TP_\filter(n \mbox{ patches}) & \leq & TP_{\model+\filter}(n+\Delta n\mbox{ patches}) \label{eq:more:patches}\\
    \time \filter(n \mbox{ patches}) & \geq & \time {\model+\filter}(n+\Delta n \mbox{ patches}) \label{eq:lower:time}
\end{eqnarray}
At least one inequality should hold in the strict sense to assure the performance of the pipeline improves with the ML model. As we shall see neither requirement is  trivially satisfied. 

Given $n$ generated patches, by definition of the prevalence rate of the generator, the good patches would be $\prevalence \cdot n$ and the true positive patches surviving the traditional validator \filter would be $\tpr \filter \cdot \prevalence \cdot n$. This process will take time $\time \filter \cdot n$. So we have defined both left terms of the two inequalities above.

After introducing the ML model \model, the pipeline asks the generator additional $\Delta n$ patches, ask \model to spend $\time \model $ to run and classify each patch, and finally re-run the validator \filter in time $\time \filter$ if the patch survives \model's screening (Fig.~\ref{fig:1b}).

The ML model will let pass to the filter the true patches $\tpr \model  \cdot \prevalence \cdot (n + \Delta n)$ and will add some false positives which depends on the precision of the ML model. Therefore the total number of patches that have to go to through the filter as specified in Figure~\ref{fig:1b} are the following ones:
\begin{align}
    TP_\model +FP_\model  = \frac{\tpr \model  \cdot \prevalence \cdot (n + \Delta n)}{\ppv \model } \label{eq:output:ML:input:f}
\end{align}

Running model \model for classifying all $n+\Delta n$ patches takes $\time \model (n+\Delta n)$ while validator \filter must run on all surviving ML patches from Eq.~\ref{eq:output:ML:input:f}. The pipeline time is therefore
\begin{align}
 \time {\model+\filter}(n+\Delta n \mbox{ patches})    = \left(\time \model  + \time F \frac{\tpr \model }{\ppv \model }\cdot \prevalence\right)  \cdot (n + \Delta n)
    \label{eq:time:f:plus:ML}
\end{align} 

We assume that the ML model will not change the distribution of the patches and thus will not change the recognition performance of the validator \filter. Thus, to compute the surviving good patches after the validator we apply its recall $\tpr \filter$ to the input that \filter receives from the ML model and namely $TP_\model$  as described in Figure~\ref{fig:1b} which is captured in Eq.~\ref{eq:output:ML:input:f}.
\begin{align}
   TP_{\model+\filter}(n+\Delta n\mbox{ patches}) = \tpr \filter \cdot \tpr \model  \cdot \prevalence \cdot (n + \Delta n)
\label{eq:tp:pipeline}
\end{align}

Replace these results into requirements~\ref{eq:more:patches} and \ref{eq:lower:time} to obtain: 
\begin{eqnarray}
    \frac{\Delta n}{n} \hspace{-1.5ex} &  \geq &  \hspace{-1.5ex} \frac{1}{\tpr \model }-1 \label{eq:bound:more:delta}\\
   \time \model  \hspace{-1.5ex} & \leq &  \hspace{-1.5ex}  \time \filter \cdot \left(\frac{n}{n + \Delta n} - \frac{\tpr \model }{\ppv \model }\cdot \prevalence\right)  \leq \time \filter \cdot \frac{\tpr \model }{\ppv \model } \cdot (\ppv \model  -\prevalence) \label{eq:bound:less:time}  
\end{eqnarray}
The first inequality (\ref{eq:bound:more:delta}) provide us the minimum number of extra patches that we must generate for the pipeline to maintain the same potentially good patches vs using the validator alone. The worse the recall, the larger the number of extra patches.

Inequality (\ref{eq:bound:less:time}) shows that the classification time of the model must decrease as the prevalence of the generated good patches increases. The precision of the model has also to be better than the ability of the generator \prevalence: if the generated patches are already enough there is no point of adding something to filter bad ones (and introducing errors into the process). Once precision is good enough, having a better recall allow to have a slower model.  Since Eq.~\ref{eq:bound:more:delta} implies $ \time \model\leq \time \filter$, it makes sense to add a ML model \emph{only if} it is faster than a traditional validator.

\subsection{Specializing to Automated Vulnerability Repair} The results above are applicable to any APR pipeline. To apply them to AVR pipelines using an ML vulnerability detector \negmodel we must solve a problem of mismatched specifications: a paper describing a ML vulnerability detector reports recall in terms of found vulnerabilities i.e. vulnerable commits and not safe patches. To \emph{use} the model, we simply reverse the classification (zeros become ones and vice versa). To \emph{compute} the performance indicators of a `positive' model \model that finds patches fixing vulnerabilities from the performance indicators of the vulnerability detector \negmodel used in reverse, we need to reverse engineer each metric. For example, the true positives $TP_\negmodel$ are vulnerable patches and they correspond to bad patches $TN_\model$ of the positive model. This is daunting as not all indicators are reported in the literature. Table~\ref{tab:models} in \S\ref{sec:experiments} shows that several papers do not report the false positive rate.

The formulae below compute the `positive' model \model parameters for Eq.~\ref{eq:bound:more:delta} and \ref{eq:bound:less:time} from the `negative' model \negmodel parameters found in the literature.
\begin{eqnarray}
\ppv\model & = & \frac{1}{1+\frac{\fpr \negmodel\cdot \ppv \negmodel \cdot (1-\tpr \negmodel)}{(1-\fpr \negmodel) \cdot (1-\ppv \negmodel) \cdot \tpr \negmodel}} \label{eq:ppv:ML:rewritten} \\
\tpr \model &  = & 1 - \fpr \negmodel 
\label{eq:tpr:ML:rewritten}
\end{eqnarray}

\section{Preliminary Experiments} \label{sec:experiments}
%
%

\subsection{Considered ML models}
At first we included 2 models for just-in-time vulnerability detection (i.e. identification of commits that potentially introduce vulnerabilities): CodeJIT RGCN and FastRGCN \cite{nguyen2024code}.  CodeJIT was trained to distinguish vulnerability-inducing and fixing commits, so although its classification on neutral commits is unpredictable, in an APR pipeline it could be eventually used to filter in the patches most similar to fixing commits.
Other models of commit predictions are not applicable: both VCCFinder \cite{perl2015vccfinder} and VulDigger \cite{yang2017vuldigger} evaluate the risk of the commit based on the committer's experience, which has no use to evaluate patches of an APR pipeline.

We further included 4 models for vulnerability detection: IVDetect \cite{li2021IVDetect}, LineVul \cite{fu2022linevul}, LineVD \cite{hin2022linevd}, VulDeePecker \cite{li2018vuldeepecker}. For VulDeePecker, we include its (poorer) performance on the Reveal dataset \cite{chakraborty2021deep}, to check how evaluating a model on different datasets can affect the effectiveness estimation. 

Table \ref{tab:models} presents the data of the selected ML 'negative' models. Most papers do not report the \fpr\negmodel. To estimate a value compatible with the collected precision \ppv\negmodel and recall \tpr\negmodel, we derive an expression of \fpr \negmodel by using Bayes’ rule $\ppv{} = (\prevalence_\negmodel \cdot \tpr{})/[\prevalence_\negmodel \cdot \tpr{} + (1-\prevalence_\negmodel) \cdot \fpr{}]$ where the prevalence $\prevalence_\negmodel$ is computed from the dataset that we can find in each paper. For example, for Vuldeepcker \cite{li2018vuldeepecker} the number of vulnerable code gadgets ($17,725$) over the  total gadgets ($61,638$) in the dataset result in a prevalence rate $\prevalence_\negmodel = 0.29$. 

\begin{table}[t]
\centering
\caption{Performance Data of candidate ML models}
\begin{minipage}{\linewidth}{\itshape \footnotesize
We gathered the data of different ML models used to detected vulnerabilities in code snippets as potential candidate for patch screeners. We collected the time to generate a prediction $\time \model $, precision $\ppv{}$, recall $\tpr{}$ and False Positive Rate $FPR$. Only one model reported the total time (starting from the raw code snippet, not the one in the dataset) needed to make a prediction.
    \vspace{0.2cm}
    }\end{minipage}
\begin{tabular}{llrrrr}
\hline
Tools          & $\time \model $ (s)& P & R & FPR & $\pi$ \\ \hline
VulDeePecker \cite{li2018vuldeepecker} & 156 & 0.87  & 0.84 & 0.05 & 0.29\\ 
VulDeePecker on Reveal \cite{chakraborty2021deep}& 156 & 0.11  & 0.14 & 0.11* & 0.09\\ 
IVDetect on Reveal  \cite{chakraborty2021deep} & $\geq 1.5$ & 0.39 & 0.52 & 0.08* & 0.09\\ 
LineVul \cite{fu2022linevul} & - & 0.97 & 0.86 & 0.002* & 0.06\\ 
LineVD \cite{hin2022linevd} & $\geq 1$ & 0.27 & 0.53 & 0.09* & 0.06\\ 
CodeJIT FastRGCN \cite{nguyen2024code}& $\geq 0.75$ & 0.77 & 0.71 & 0.22 & 0.5\\ 
CodeJIT RGCN \cite{nguyen2024code}& $\geq 1.42$ & 0.78 & 0.70 & 0.20* & 0.5\\ \hline
\end{tabular}
\begin{minipage}{\linewidth}{\itshape \footnotesize
*estimated through Bayes’ rule $\ppv{} = (\prevalence \cdot \tpr{})/[\prevalence \cdot \tpr{} + (1-\prevalence) \cdot \fpr{}]$
    \vspace{0.2cm}
    }\end{minipage}
\label{tab:models}
\end{table}

A key, severely under-reported issue is the time of the model. Most papers mention the time to query the models\emph{only after} the code has been transformed into a ML ingestible format. Only the authors of VulDeePecker included the time to pre-process source code into an ML format. This time is important for a fair comparison as the validator \filter starts from the source code. Since the transformation is different from paper to paper, we express the models' potential effectiveness by giving upper bounds on their total potential pre-processing and query time.

\subsection{Estimation}
\paragraph{Fixed ML model} given an ML model, we estimate the characteristics of an AVR validation process that the model could improve. The VulDeePecker \cite{li2018vuldeepecker} model trained on the HY-ALL dataset is the only one provided with all the values for our estimation. However, the model performance was quite different when tested on ReVeal, a dataset of real-world vulnerabilities \cite{chakraborty2021deep}. The data is reported in Table~\ref{tab:time}.
Using equations~ \ref{eq:bound:more:delta} and \ref{eq:bound:less:time} with the given values, introducing the original VulDeePecker model before a validation filter $\filter$is convenient if the filter time is $\time {\filter, original} > 4.56min$ and $\Delta n / n _{\filter, original} > 5.26\%$ more candidate patches are added. While using the data on the Reveal dataset the minimum requirements are $\Delta n / n _{\filter, ReVeal} > 12.1\%$ and $\time {\filter, ReVeal} > 5.07min$.
\paragraph{Fixed AVR pipeline} given an AVR pipeline, we estimate the prediction time limits for ML models to improve its validation process. We considered the framework of APR4Vul \cite{bui2024apr4vul}. In particular, we used the measured ratio of correct patches ($30$) over the total generated patches ($78$) to estimate a realistic prevalence of safe patches $\pi=0.38$, and the testing times for the Vul4J \cite{bui2022vul4j} benchmark of Java vulnerabilities. 
Each vulnerability in Vul4J is provided with both unit tests to be used as functional requirements for the repair patch and Proof-of-Vulnerability tests to verify the vulnerability presence.
The authors indicate that the full test suite requires less than $9.17s$ for the fastest quartile of vulnerabilities, less than $27.04s$ for the second quartile, and less than $74.5s$ for the third quartile. 
However, the mean test execution time is $337.83s$ for the presence of some outliers. 

These time values are used as reference filter time $\time {\filter}$ one for each column of Table \ref{tab:time}. The table reports the time limit for the ML validation of a single patch $\time {ML}$ (including pre-processing) for the ML model to act as an effective pre-screener before testing. The time limits are computed using Eq.~\ref{eq:bound:less:time} filled with the equivalent precision $\ppv\model$ (from Eq.~\ref{eq:ppv:ML:rewritten}) and recall $\tpr\model$ (from Eq.~\ref{eq:tpr:ML:rewritten}) of each vulnerability-detection model, the prevalence rate $\prevalence=0.38$ of correct patches obtained from APR4Vul \cite{bui2024apr4vul} and the filter time $\time {\filter}$ corresponding to each column.

\finding{\textbf{\textit{Preliminary result.}} 
Even the most effective model would need less than $5.67s$ to classify a patch of interest to be a convenient pre-validator when the unit tests of APR4Vul are in the quickest 25\%. As the only indicated execution time with pre-processing is over 2.5 minutes (VulDeePecker), current models may not be effective in filtering most of the unpromising patches before testing.}

\begin{table}[t]

\centering
\caption{Time limit for predicting while still being an effective pre-screener}
\begin{minipage}{\linewidth}{\itshape \footnotesize
We compute the maximum pre-processing and prediction time $T_\model $ that a model could use to be an effective pre-screener before testing. The first column reports the time limit for $T_\model $ for the introduction of the models to be convenient before 75\% of the slowest test suites, the second for 50\%, and the third for the last 25\%. Due to the presence of large outliers, the mean $T_\filter$ is not a reliable indicator for the prediction time limit. All times in seconds.
    \vspace{0.2cm}
    }\end{minipage}
\begin{minipage}{\linewidth}{\itshape \footnotesize

    \vspace{0.1cm}
    }\end{minipage}
\begin{tabular}{lrrrr}
\hline
Tools & Q25 & Median & Q75 & Mean
\\ \hline
VulDeePecker & 5.23 & 15.6  & 42.5 & 193 \\ 
VulDeePecker on ReVeal & 4.70 & 14.0  & 38.2 & 173  \\ 
IVDetect on ReVeal & 4.95 & 14.8  & 40.2 & 182 \\ 
LineVul & 5.67 & 16.9  & 46.1 & 209 \\ 
LineVD & 4.85 & 14.5 & 39.4 & 179 \\ 
CodeJIT FastRGCN & 3.67 & 11.0 & 29.8 & 135 \\ 
CodeJIT RGCN & 3.87 & 11.6 & 31.5 & 143 \\ \hline
\end{tabular}
\label{tab:time}
\end{table}


\section{Future plans and limitations}
\label{sec:future:plans}

In this paper, we propose to break down the validation process of APR patches into steps to make it more efficient. 
We envision the first step to be an ML model performing binary classification to quickly discard most of the unpromising patches before they undergo a second, more expensive validation filter based on testing.
We use the model, filter and data distribution properties to provide an estimation of whether introducing a model before a testing pipeline could improve the validation process. 
The preliminary experiments seems to confirm the potentiality of our results but a conducting a detailed experimental and theoretical analysis is needed. 

\textbf{Systematic setup.} The data collected from the literature helps us estimate whether it \textit{might} be useful to introduce an ML model. Still, several papers only report partial data (see Table~\ref{tab:models}) and can vary based on external factors (e.g. the datasets). 
Additionally, while we gathered data on the execution of full unit tests, approaches that only run subsets of test cases per iteration might lead to differing run times per validation cycle. 
We aim to select a baseline AVR pipeline and a systematic setup to empirically measure the ML models' contribution to its validation process.  

\textbf{ML models inclusion.} We want to expand our evaluation to other ML models, either trained to distinguish vulnerability-fixing commits or predict whether the repair patch is compilable.

\textbf{Generating more patches.} Finally, we want to verify whether ranting a repair patch generator the time to produce more patches could \textit{actually} improve the chances of selecting good repairs. Bui et al. \cite{bui2024apr4vul} observed how APR test-based tools often do not generate patches to repair vulnerabilities, thus improving the available time may not impact their results \cite{vu2021please}. However, granting an ML-based patch generator more attempts could yield better result, if the generation is not too slow.


\section{Acknowledgements}

This work has been partly supported by the European Union (EU) under Horizon Europe grant n.\ 101120393 (Sec4AI4Sec), by the Italian Ministry of University and Research (MUR), under the P.N.R.R. – NextGenerationEU grant n.\ PE00000014 (SERICS), and by the Nederlandse Organisatie voor Wetenschappelijk Onderzoek (NWO) under grant n.\ KIC1.VE01.20.004 (HEWSTI).

\section{CRediT Author Statement}
Conceptualization MC, FM;
Methodology MC, FM; 
Validation MC, FM;
Formal analysis FM, MC;
Investigation MC;
Data Curation MC; 
Writing - Original Draft MC;
Writing - Review \& Editing MC, FM;
Visualization MC;
Supervision FM;
Project administration FM;
Funding acquisition FM.

\IEEEtriggeratref{15}
\bibliographystyle{IEEEtran}
\bibliography{references}

\clearpage

\section*{Appendix - Formal Derivations}

\subsection{Formal derivations}

We consider the formal requirements expressed in  
Eq.~\ref{eq:more:patches} and  Eq.~\ref{eq:lower:time} (and reported here for convenience) for an ML model to be an effective pre-screener in the validation process of an APR pipeline.

\begin{eqnarray*}
    TP_\filter(n \mbox{ patches}) & \leq & TP_{\model+\filter}(n+\Delta n\mbox{ patches}) \hspace{7ex} (\ref{eq:more:patches})\\
    \time \filter(n \mbox{ patches}) & \geq & \time {\model+\filter}(n+\Delta n \mbox{ patches}) \hspace{9ex} (\ref{eq:lower:time})
\end{eqnarray*}

Given $n$ patches as input, the original validator will produce the positive patches below
\begin{align}
TP_\filter(n \mbox{ patches}) = \tpr \filter \cdot \prevalence \cdot n
\label{eq:TP:F}
\end{align}

Before the introduction of the model, the time required by the APR pipeline to validate $\totalpatches$ patches is just the time to run the validator $\time F$ on all candidate patches $\totalpatches$ and namely
\begin{align}
\time \filter(n \mbox{ patches})  = \time \filter \cdot n 
\label{eq:time:f}\end{align}

After introducing the ML model, the pipeline will ask the generator to generate additional $\Delta \totalpatches$ patches and ask the ML model to spend $\time \model $ to run and classify each patch, and finally re-run the filter with time $\time \filter$ on the ML-surviving patches.

Running the model would therefore cost at least the amount of time below:
\begin{align}
    \time \model (\totalpatches+\Delta \totalpatches) \label{eq:time:ML}
\end{align} 

Then, we have to filter the resulting patches, which are classified by the ML model as positive, through the filter. 
So the ML model will pass to the filter the following true patches
\begin{align}
    TP_\model  = \tpr \model  \cdot \prevalence \cdot (\totalpatches + \Delta \totalpatches)
    \label{eq:TP:ML}
\end{align}
and will add some false positives which depends on the precision of the ML model which might be potentially fewer than the initial wrong patches $(1-\prevalence) \cdot \totalpatches$ supplied by the generator. 

The precision of the model will determine the relative ratio of the true positive and false positives, thus the total number of patches that have to go to through the filter as specified in Figure~\ref{fig:1b} are expressed by Eq.~\ref{eq:output:ML:input:f} (reported here for convenience)
\begin{align*}
    TP_\model +FP_\model  = \frac{\tpr \model  \cdot \prevalence \cdot (n + \Delta n)}{\ppv \model } \hspace{20ex}(\ref{eq:output:ML:input:f})
\end{align*}

This means that the filter will have to run for at least the following time
\begin{align}
    \time \filter \cdot \frac{\tpr \model  \cdot \prevalence \cdot (n + \Delta n)}{\ppv \model }
    \label{eq:time:f:afterML}
\end{align} 

We can now compute the overall time required by the pipeline by combining the time for running the model from Eq.~\ref{eq:time:ML} and the time for running the filter after the ML model from Eq.~\ref{eq:time:f:afterML}. 
\begin{align}
    \time \model (n+\Delta n) + \time \filter \cdot \frac{\tpr \model  \cdot \prevalence \cdot (n + \Delta n)}{ \ppv \model }
\end{align}
By reorganizing the terms, we obtain the desired time to run the overall pipeline on the original plus additional patches in Eq.~\ref{eq:time:f:plus:ML}, here reported for convenience.
\begin{align*}
 \time {\model+\filter}(n+\Delta n \mbox{ patches})    = \left(\time \model  + \time \filter \frac{\tpr \model }{\ppv \model }\cdot \prevalence\right)  \cdot (n + \Delta n) 
 ~(\ref{eq:time:f:plus:ML})
\end{align*} 

We assume that the ML model will not change the distribution of the patches and thus will not change the recognition performance performance the filter. Therefore, to compute the surviving good patches after the filter we need to apply the recall of the filter $\tpr F$ to the true positive input 
that it receives from the ML model and namely $TP_\model$  as described in Figure~\ref{fig:1b}. Using the $TP_\model$ definition from Eq.~\ref{eq:TP:ML}, we obtain Eq.~\ref{eq:tp:pipeline}, reported here for convenience.
\begin{align*}
   TP_{\model+\filter}(n+\Delta n\mbox{ patches}) = \tpr \filter \cdot \tpr \model  \cdot \prevalence \cdot (n + \Delta n)
~(\ref{eq:tp:pipeline})
\end{align*}

We can now replace the obtained results from equations (\ref{eq:TP:F}, \ref{eq:time:f},  \ref{eq:time:f:plus:ML}, \ref{eq:tp:pipeline}) into our requirements inequalities Eq.~\ref{eq:more:patches} and Eq.~\ref{eq:lower:time} as follows:
\begin{align}
     \tpr \filter \cdot \prevalence \cdot n \leq  \tpr \filter \cdot  \tpr \model  \cdot \prevalence \cdot (n + \Delta n) \label{eq:more:patches:expanded}\\
     \time \filter \cdot n \geq  \left(\time \model  + \time \filter \frac{\tpr \model }{\ppv \model }\cdot \prevalence\right)  \cdot (n + \Delta n) \label{eq:lower:time:expanded}
\end{align}

We can now simplify by canceling $ \tpr F \cdot \prevalence $ from both sides of the first inequality to obtain
\begin{align}
     \frac{n}{n + \Delta n} \leq \tpr \model  \label{eq:bound:patches}
\end{align}
For the second inequality we obtain
\begin{align}
     \time F \cdot \frac{n}{n + \Delta n} \geq  \time \model  + \time F \frac{\tpr \model }{\ppv \model }\cdot \prevalence  
\end{align}
and by moving $\time \model $ on the left and swapping the sign we obtain
\begin{align}
   \time \model   \leq \time F \cdot \left(\frac{n}{n + \Delta n} - \frac{\tpr \model }{\ppv \model }\cdot \prevalence\right)
   \label{eq:bound:time}
\end{align}
By refactoring Eq.~\ref{eq:bound:patches}, we obtain Eq.~\ref{eq:bound:more:delta}, which expresses the minimum ratio of extra patches that we must generate for the pipeline to maintain the same potentially good patches vs using the validator alone. 

Eq.~\ref{eq:bound:patches} provides an upper bound on the ratio between $n$ and $n+\Delta n$. We replace its the left-hand in the right side of Eq.~\ref{eq:bound:time} to maximize the bound on the time, and by factoring $\tpr \model $ as a common term, we obtain the
desired upper bound on the classification time of the ML algorithm in Eq.~\ref{eq:bound:less:time}. We report here equations Eq.~\ref{eq:bound:more:delta} and Eq.~\ref{eq:bound:less:time} for convenience.
\begin{eqnarray*}
     \frac{\Delta n}{n} & \geq & \frac{1}{\tpr \model }-1 \hspace{35ex}(\ref{eq:bound:more:delta})\\
   \time \model  & \leq & \time \filter \cdot \frac{\tpr \model }{\ppv \model } \cdot (\ppv \model  -\prevalence) \hspace{25.5ex}(\ref{eq:bound:less:time})
\end{eqnarray*}

\subsection{Discarding unpromising patches.} 
The results from Eq.~\ref{eq:bound:more:delta} and Eq.~\ref{eq:bound:less:time} are applicable to any APR pipeline and any ML models detecting safe patches. To apply them using the performance data of an ML model trained to detect vulnerable patches, we need to reverse engineer each metric. By reversing the
classification, we obtain that true positives become true negatives, false positives become false negatives, and so on.
\begin{align}
TP_\negmodel = TN_\model  \label{eq:TP:MLminus:TNML} \\
FP_\negmodel = FN_\model   \label{eq:FP:MLminus:FNML} \\
FN_\negmodel = FP_\model  \label{eq:FN:MLminus:FPML} \\
TN_\negmodel = TP_\model   \label{eq:TN:MLminus:TPML}
\end{align}
The formal definition of Precision, Recall and False Positive Rate of model detecting vulnerabilities are the following ones
\begin{align}
\ppv \negmodel = \frac{TP_\negmodel}{TP_\negmodel+FP_\negmodel} \label{eq:ppv:ML:minus} \\
\tpr \negmodel = \frac{TP_\negmodel}{TP_\negmodel+FN_\negmodel} \label{eq:tpr:ML:minus} \\
FPR_\negmodel = \frac{FP_\negmodel}{FP_\negmodel+TN_\negmodel} \label{eq:fpr:ML:minus}
\end{align}
We now replace in the definitions above the I/O mapping defined in the equations (\ref{eq:TP:MLminus:TNML}, \ref{eq:FP:MLminus:FNML}, \ref{eq:FN:MLminus:FPML},
\ref{eq:TN:MLminus:TPML}). So we replace $TP_\negmodel$ in Eq.~\ref{eq:ppv:ML:minus} with $TN_\model $ from Eq.~\ref{eq:TP:MLminus:TNML} and $FP_\negmodel$ with $FN_\model $ from Eq.~\ref{eq:FP:MLminus:FNML} to obtain a new definition for $\ppv \negmodel$ which is presented in the Eq.~\ref{eq:ppv:ML:minus:rewritten}. The same process is repeated for Eq. \ref{eq:tpr:ML:minus} yielding \ref{eq:tpr:ML:minus:rewritten} and to Eq.~\ref{eq:fpr:ML:minus} yielding \ref{eq:fpr:ML:minus:rewritten}.
\begin{align}
\ppv \negmodel = \frac{TN_\model }{TN_\model +FN_\model } \label{eq:ppv:ML:minus:rewritten} \\
\tpr \negmodel = \frac{TN_\model }{TN_\model +FP_\model }   \label{eq:tpr:ML:minus:rewritten} \\
FPR_\negmodel = \frac{FN_\model }{FN_\model +TP_\model }  \label{eq:fpr:ML:minus:rewritten}
\end{align}

Eq.~\ref{eq:ppv:ML:minus:rewritten} can be used to express $FN_\model$ in terms of $TN_\model $ in Eq. \ref{eq:FNML:expanded}, 
Eq.~\ref{eq:tpr:ML:minus:rewritten} to express $FP_\model$ in terms of $TN_\model $ in Eq. \ref{eq:FPML:expanded} and Eq.~\ref{eq:fpr:ML:minus:rewritten} to express $TP_\model$ in terms of $FN_\model $ in Eq. \ref{eq:TPML:expanded}.

\begin{align}
FN_\model  = \left(\frac{1}{\ppv \negmodel}-1\right)\cdot TN_\model  \label{eq:FNML:expanded} \\ 
FP_\model  = \left(\frac{1}{\tpr \negmodel}-1\right)\cdot TN_\model  \label{eq:FPML:expanded} \\
TP_\model  = \left(\frac{1}{FPR_\negmodel}-1\right)\cdot FN_\model  \label{eq:TPML:expanded}
\end{align}

We can combine equations~\ref{eq:FNML:expanded} and \ref{eq:TPML:expanded} to express $TP_\model $ in terms of $TN_\model $
\begin{align}
TP_\model  = \left(\frac{1}{FPR_ \negmodel}-1\right)\cdot \left(\frac{1}{\ppv \negmodel}-1\right)\cdot TN_\model  \label{eq:TPML:expanded:2}
\end{align}

We can use these equations to obtain the equivalent precision of the model if it was to select safe patches. We start with the formal definition of precision in Eq. \ref{eq:ppv:ML}
\begin{align}
\ppv\model  = \frac{TP_\model }{TP_\model +FP_\model } \label{eq:ppv:ML}
\end{align}
We expand the righthand term with the characterization of $FP_\model $ that we have computed in Eq.~\ref{eq:FPML:expanded} and the expression of $TP_\model $ that we have obtained from Eq.~\ref{eq:TPML:expanded:2}.
\begin{align}
\frac{\left(\frac{1}{FPR_\negmodel}-1\right)\cdot \left(\frac{1}{\ppv \negmodel}-1\right)\cdot TN_\model }{\left(\frac{1}{FPR_\negmodel}-1\right)\cdot \left(\frac{1}{\ppv \negmodel}-1\right)\cdot TN_\model +\left(\frac{1}{\tpr \negmodel}-1\right)\cdot TN_\model } \nonumber
\end{align}
We perform some algebraic transformations (e.g. we cancel $TN_\model $ in the numerator and the denominator) and obtain the following form for the right-hand term of Eq.~\ref{eq:ppv:ML}
\begin{align} 
\frac{1}{1+\frac{\left(\frac{1}{\tpr \negmodel}-1\right)}{\left(\frac{1}{FPR_\negmodel}-1\right)\left(\frac{1}{\ppv \negmodel}-1\right)}}\nonumber
\end{align}
This finally yield Eq.~\ref{eq:ppv:ML:rewritten}, reported below for convenience, in which 
the precision for the detection of safe patches $\ppv\model $ is been expressed in terms of the precision $\ppv \negmodel$, recall $\tpr \negmodel$ and false positive rate $FPR_\negmodel$ of the model for detecting vulnerabilities. 
\begin{align*}
\ppv\model  = \frac{1}{1+\frac{FPR_\negmodel\cdot \ppv \negmodel \cdot (1-\tpr \negmodel)}{(1-FPR_\negmodel) \cdot (1-\ppv \negmodel) \cdot \tpr \negmodel}} \hspace{19ex}(\ref{eq:ppv:ML:rewritten})
\end{align*}

Eq. ~\ref{eq:fpr:ML:minus:rewritten} can be used to rewrite the formal definition of the recall of the model detecting safe patches $\tpr \model$ in terms of the recall of the vulnerability detection model $FPR_\negmodel$. The result is Eq.~\ref{eq:tpr:ML:rewritten}, reported here for convenience.
\begin{align*}
\tpr \model  = \frac{TP_\model }{TP_\model +FN_\model } = 1 - FPR_\negmodel 
\hspace{19ex}(\ref{eq:tpr:ML:rewritten})
\end{align*}

\end{document}